\documentclass[11pt,a4paper]{article}
\pdfoutput=1

\usepackage{jheppub}
\usepackage{amssymb}
\usepackage[T1]{fontenc}
\usepackage[utf8]{inputenc}
\usepackage[makeroom]{cancel}
\usepackage{graphicx}

\usepackage{physics}
\usepackage{amsmath,amssymb,amsthm,textcomp}
\usepackage{mathrsfs}
\usepackage{enumerate}
\usepackage{tikz}
\usepackage{hyperref}
\usepackage{bbm}
\usepackage{mathtools}
\usepackage{standalone}

\usepackage{bm}
\usepackage{natbib}
\usepackage{tensor}

\newcommand{\innp}[2]{\langle{#1},{#2}\rangle}

\usepackage{stackengine}
\stackMath
\renewcommand{\vec}[1]{\mathbf{#1}}
\newcommand{\vk}{\vec{k}}

\newcommand{\vA}{\vec{A}}

\title{Unitary quantization of a scalar charged field and Schwinger effect}
\author[a,b]{Luis J. Garay,}
\author[a,c]{Alberto Garc\'{\i}a Mart\'{\i}n-Caro}
\author[a]{and Mercedes Mart\'{\i}n-Benito}
\affiliation[a]{Departamento de F\'{\i}sica Te\'orica and  IPARCOS, Universidad Complutense 
de Madrid, Plaza de las Ciencias 1, 28040 Madrid, Spain}
\affiliation[b]{Instituto de Estructura de la Materia (IEM-CSIC), Serrano 121, 28006 Madrid, Spain}
\affiliation[c]{Departamento  de F\'{\i}sica  de  Part\'{\i}culas,  Universidad  de  Santiago  de  Compostela  and  Instituto Galego  de  F\'{\i}sica  de  Altas  Enerxias  (IGFAE), E-15782 Santiago  de  Compostela,  Spain}
\emailAdd{luisj.garay@ucm.es}
\emailAdd{alberto.martin-caro@usc.es}
\emailAdd{m.martin.benito@ucm.es}
\abstract{
Quantum field theory in curved spacetimes suffers in general from an infinite ambiguity in the choice of Fock representation and associated vacuum. 
In cosmological backgrounds, the requirement of a unitary implementation of the field dynamics in  the physical Hilbert space of the theory is a good criterion to ameliorate such ambiguity. Indeed, this criterion, together with a unitary implementation of the symmetries of the equations of motion, leads to an equivalence class of unitarily equivalent quantizations that, even though it is still formed by an infinite number of Fock representations, is unique.
In this work, we apply the procedure developed for fields in cosmological settings to analyze the quantization of a scalar  field in the presence of an external electromagnetic classical field in a flat background.  We find a natural Fock representation that admits a unitary implementation of the quantum field dynamics. It automatically allows to define a particle number density at all times in the evolution with the correct asymptotic behavior, when the electric field vanishes.  Moreover we show the unitary equivalence of all the quantizations that fulfill our criteria, so that they form a unique equivalence class. Although we perform the field quantization in a specific gauge, we also show the equivalence between the procedures taken in different gauges. }
\arxivnumber{1234.5678}

\begin{document}

\maketitle
\flushbottom 

\section{Introduction}
The process of spontaneous creation of particle-antiparticle pairs due to the presence of strong electro\-magnetic fields is usually known as Schwinger effect.  This effect was proposed by F. Sauter \cite{Sauter1931742}, and first studied by means of the effective action of a charged particle in an external strong electromagnetic background by Heisenberg and Euler \cite{Heisenberg1936} and Weisskopf \cite{Weisskopf:1996bu}. Finally, Schwinger gave the explanation of this effect in the full quantum electrodynamics theory \cite{Schwinger1951} for the case of a constant field. Since these seminal works, the Schwinger effect has been extensively treated in the literature by means of many different approaches, both from the theoretical and experimental points of view. Most of the previous theoretical works focus on the calculation of the particle number density, which has been obtained for simple external field configurations via the effective action method pioneered by Heisenberg and Euler  \cite{DUNNE_effeclag}, the Wigner function formalism \cite{Schwingerwigner,Hebenstreit:2010cc}, the quantum Vlasov equation \cite{Kluger:1998bm, Fedotov2011,Huet2014} and the canonical quantization approach \cite{Gavrilov:1996pz,Pady2019okd}. See also \cite{Gelis_2016} for a recent review on these and other methods. In this work we will address the Schwinger effect  within the  canonical quantization  approach, making special emphasis in the problem of unitary implementation of the evolution.  

From the experimental point of view, strong-field quantum effects of the QED vacuum will be important for electric field intensities of the order of magnitude of the QED critical field, $E^*=m^2/e\approx 1.32\times10^{18}\text{V}/\text{m}$, also called Schwinger limit \cite{Yakimenko:2018kih}. Due to the extreme technical difficulty of creating such strong electric fields, this effect, as well as other strong-field QED effects such as diphoton conversion to electron-positron pairs (the so called Breit-Wheeler process, \cite{Brewheel}), have not been yet observed. However, it is conceivable to being able to measure electron-positron pair production from the QED vacuum at planned facilities \cite{Alkofer:2001ik}, and there has been an increasing interest about this topic in the literature in recent years, due also to the fact that both theoretical and experimental ideas for overpassing and/or reducing the field intensity threshold, such as the dynamical assistance \cite{dinamicallyassisted},  have been proposed. See also \cite{Yakimenko:2018kih,Magnusson:2019qop} and references therein for recent experimental proposals to explore the strong-field QED, as well as the state of the art of high intensity laser facilities and other experiments expected to operate in regimes where these strong field effects become important. 

Similar processes of spontaneous pair creation also appear in the theory of quantum fields in curved spacetimes \cite{BlackHexplosons,LParker,GibbHawk}. Furthermore, there are several common features in both types of processes, such that they could be gathered into a broader framework, namely the quantum field theory in non\-trivial (classical) background fields, where the nature of this background may be an external gauge field, the curvature of spacetime, or both at the same time, and plays an important role in the evolution of the system.

The most common way of describing particle creation by an external field involves the asymptotic analysis of particle states both in the distant past (before the external field is `switched on') and future (long after the interaction with the external field has finished). The quantum field operator is assumed to reduce to an expression in terms of creation and annihilation operators, associated to one-particle states both in the past and the future. Then, from the relation between both sets of operators in the past and the future we can compute an expression for the $S$-matrix of the process, and the number of particles created in the process. This is the approach followed, for example, in \cite{WALD1979490,Gavrilov:1996pz}.

This approach, however, is not completely satisfactory for several reasons. Firstly, since it only gives information about the asymptotic states of the quantum field, it cannot give a complete description of the process when the quantum field and the external background are still interacting.  In particular, there exists an ambiguous definition of the density of created particles at intermediate times \cite{Dabrowski_2016,Gelis_2016, Hebenstreit:2008ae, Kim:2011sf}. Also, this approach is based on the concept of asymptotic free-particle states, which, as we have already mentioned, in general  cannot be uniquely defined in a nontrivial background, and may not even exist \cite{wald1994quantum}. 

Indeed, in these kind of systems, the lack of symmetry of the underlying background (compared to empty Minkowski spacetime) implies the existence of several ambiguities in the quantization procedure of the classical field. In particular, one of the problems that must be dealt with when quantizing a field in a non-trivial background is the infinite ambiguity that arises whenever one tries to choose a particular representation of the creation and annihilation operators on the Hilbert space of solutions (Fock representation),  since  there usually exist different, non-equivalent Fock representations. These non-equivalent representations lead to different notions of the concept of particle (or, equivalently, different definitions of the vacuum state of the theory). This is not a problem in Minkowski spacetime, since  Poincaré invariance of the vacuum state is usually imposed, and this condition uniquely defines a family of basis sets of modes which are related via unitary transformations (these transformations being the Poincaré transformations represented as unitary operators). However, in non-trivial backgrounds, symmetry conditions are often not sufficient to reduce the ambiguity in the choice of vacuum states, and some other conditions must be imposed in order to obtain a unique quantization procedure.

In this work, we take advantage of the great similarity between the Schwinger pair creation effect and other pair creation processes due to the curvature of spacetime, and approach it with the mathematical tools found in the treatment of the latter. In particular,  in the context of quantum fields in cosmological spacetimes a criterion has been put forward to  drastically reduce  this ambiguity, which consists in imposing the unitary implementation of the evolution at all times, as succinctly explained in \cite{Cortez:2015mja, Cortez_2020}. 
This criterion is well motivated, since time evolution in quantum mechanics is characterized by unitary operators.
Moreover, together with a unitary implementation of the symmetries, it serves to select a unique  family of unitarily equivalent Fock representations  in a diverse variety of cosmological settings, such as isotropic spacetimes with both scalar fields (see e.g. \cite{Gomar:2012xn, Cortez:2012cf}) and fermionic fields (see e.g. \cite{Cortez:2016kwy,Cortez:2016xsn}), or anisotropic Bianchi I spacetimes with a scalar field \cite{Cortez:2016oxm}.
By imposing this condition, one forces not only the initial and final states of the field (asymptotic states) to be related by a unitary transformation given by the asymptotic $S$-matrix, (whose existence was proven in \cite{WALD1979490} for general backgrounds) but also that the intermediate states of the field must be related through a unitary operator, namely, the evolution operator.

Following the same kind of reasoning, we propose this criterion of unitary quantum dynamics for the case of a charged scalar field. The problem of the unitary implementation of the evolution of a quantum field in an arbitrary external electromagnetic field has been treated in the literature before \cite{Ruijsencharged,thaller1992dirac}. These previous works suggest that it is not possible, in general, to unitarily implement the evolution into the Hilbert space defined by the Minkowski vacuum whenever an arbitrary external electromagnetic potential is present, unless the external potential satisfies some rather restrictive conditions. 
However, as we show in this work, unitary implementation of the evolution can be achieved for a wider class of external potentials by choosing an appropriate set of creation and annihilation variables to quantize the theory. From a mathematical point of view, this means that one can take advantage of the   freedom in the choice of Fock representations of the classical theory, (equivalently, the choice of a complex structure which defines the set of creation and annihilation variables) by choosing one that allows the unitary implementation of the dynamics. 

The main result of our study is the proof that all Fock representations that allow for a unitary quantum implementation of the dynamics (together with a unitary quantum implementation of the symmetries of the equation of motion) belong to a unique family of unitarily equivalent quantizations. Thanks to the unitarity of the quantum dynamics, all of them lead to a  particle number density of created particles via Schwinger effect that is well defined at all times in the evolution. From this perspective, our study complements previous works that analysed the creation of particles by Schwinger effect, as those of \cite{Kluger:1998bm, Huet2014,Fedotov2011} which are based on a concrete choice of quantization characterized by a basis of adiabatic states, or the analysis of the time evolution of the number operator done in \cite{Dabrowski_2016}, whose basis choice is associated with a particular truncation of the adiabatic expansion. We conclude that any quantization with a well-defined expectation value of the particle number density along the whole evolution (as the ones mentioned) must belong to our unique equivalence class.

We also show that the gauge freedom in the description of the external field can be consistently taken into account. In particular, the presented procedure ensures that quantizations carried out in different gauges yield the same physical predictions, i.e. the same particle creation.

The structure of this paper is the following. In section II, we review the classical theory of a complex scalar field and its canonical quantization, emphasizing the important role played by the complex structure in this procedure.  Section III is devoted to the analysis of canonical transformations of the classical field variables (the so-called Bogoliubov transformations), and to establish under which conditions these transformations can be implemented as unitary operators into the Fock space of the quantum theory. As a special case, we   regard the classical evolution as a Bogoliubov transformation, and study the restrictions imposed on the complex structure by the requirement of unitary implementation of this transformations. We also analyze the gauge freedom in the description of the external electromagnetic fields.
In section IV,  we particularize our analysis to the case of a spatially homogeneous electric field with arbitrary time dependence. We show that the unitary
implementation of the quantum field evolution reduces the ambiguity on the choice of complex structure up to unitary
equivalence. This is achieved by adapting the analyses of \cite{Cortez:2016oxm,Cortez:2016xsn,Cortez:2016kwy} to our system.
Finally, in section V, we provide an explicit example of the application of the procedure developed in the previous sections to the scalar Schwinger effect in a concrete external potential, the Sauter pulse \cite{Sauter1931742}.  In particular, we show explicitly how our method allows to define a particle number density of created particles via Schwinger effect at any instant of time, which agrees with the one obtained in previous works using the quantum Vlasov equation \cite{Huet2014}, and also presents the same asymptotic limit as the one obtained in \cite{Gavrilov:1996pz,Navarro} by asymptotic approaches.
We conclude in section VI with a recap of the main results achieved in this paper.

We use units such that $c=\hbar=1$.

 \section{Canonical quantization of a charged scalar field in an electromagnetic background}

\label{classicalscalar}
Let us consider  a charged scalar field $\phi$ coupled to an electromagnetic field specified by the four-vector potential $A_\mu$, in flat spacetime. Its action is given by
\begin{align}
   S=-\int d^4x\big[(\partial^\mu-iqA^\mu){\phi^*} (\partial_\mu+iqA_\mu)\phi+m^2\phi^*\phi\big],
   \label{scalaraction}
\end{align}
where $q$ is the electric charge of the charged field, $m$ is its mass, and the symbol $*$ denotes complex conjugation.  The equation of motion for this scalar field can be variationally obtained from this action:
\begin{align}
    [(\partial^\mu+iqA^\mu) (\partial_\mu+iqA_\mu)+m^2]\phi(x)=0.
     \label{ChargedKG}
\end{align}

The canonical phase space of the theory $\Gamma$ is understood as the Cauchy data space (or space of initial conditions) endowed with the corresponding Poisson bracket structure. An alternative description of the phase space of the theory can be given in terms of the (covariant) space of solutions $\mathcal{S}$ of the  equation of motion \eqref{ChargedKG}.
As discussed in \cite{wald1994quantum}, for a linear field there is a one-to-one correspondence between elements of $\Gamma$ and those of $\mathcal{S}$. In other words, the classical theory can be described both by pairs of fields in a Cauchy hypersurface $\Sigma$ (canonical approach) or by solutions of \eqref{ChargedKG} (covariant formalism).

Equation \eqref{ChargedKG} is gauge covariant, in the sense that if $\phi$ satisfies \eqref{ChargedKG}, then $e^{iqf(x)}\phi$ satisfies the same equation, with $A_\mu$ replaced by $A_\mu-\partial_\mu f(x)$. Indeed, since we are considering that the gauge field is an external (nondynamical) field, the local gauge symmetry of the action does not imply that the equations of motion ---understood as equations for the matter fields only--- are invariant (as would be the case for global gauge symmetries), but covariant under that symmetry in the sense commented above. Thus, gauge transformations in this sense can be understood as (possibly time-dependent) canonical transformations of the field variables that may change the explicit form of the equations of motion.
Note that, since the explicit form of the equations of motion depends on the particular electromagnetic potential chosen, the construction of this space will also depend on that choice. Furthermore, the full covariant phase space will be gauge dependent (although the relation between the elements of two spaces of solutions corresponding to two gauge-related potentials is one-to-one, given by a gauge transformation). Physical results should be the same when describing the system by two gauge-related covariant phase spaces, since no physically measurable quantity should depend on the particular gauge chosen to calculate it. We will come back to this issue in section \ref{gauge}.

As shown in \cite{Lee:1990nz}, the covariant phase space $\mathcal{S}$ admits a symplectic structure $\Omega$ constructed from the Lagrangian density. In the case that we are studying, the symplectic form $\Omega$ can be written as 
\begin{align}
      \Omega( \phi, \psi)=\Re\big[\mu(\phi,\psi)\big],
\label{symplecscalar}
\end{align}
for each pair of solutions $\phi$ and $\psi$, where $\mu$ is the sesquilinear form on $\mathcal{S}$ 
\begin{align}
        \mu(\phi,\psi)=\int_{\Sigma} d^3\vec{x}\big[\psi^*(\partial_0+iqA_0)\phi-\phi(\partial_0-iqA_0)\psi^*\big].
\label{mu}
\end{align} 
Note that 
\eqref{mu} is preserved both by the evolution and gauge transformations and hence, so is $\Omega$.

In order to quantize the classical theory, we need to define a complex Hilbert space constructed from the linear space of complex solutions $\mathcal{S}$ by means of a (time and gauge-independent) Hermitian inner product.  Note that the application $\mu$ defined in \eqref{mu} satisfies these requirements, but fails to be positive definite. We may thus define an Hermitian inner product by introducing a complex structure $J$ acting on $\mathcal{S}$ such that
\begin{align}
    \ev{\cdot,\cdot}\equiv\mu(\cdot,J\cdot)
    \label{innerprod}
\end{align}
is a positive definite inner product in $\mathcal{S}$.  This also ensures that such $J$ is compatible with the symplectic structure, i.e. that $\Omega(J\cdot,J\cdot)=\Omega(\cdot,\cdot)$.
Now let 
\begin{align}
    P_J^\pm=\frac{1}{2}(\mathbbm{1}\mp iJ)
    \end{align} 
be the projector operators on the spectral eigenspaces of $J$ in $\mathcal{S}$, i.e. $JP_J^\pm \mathcal{S}=\pm i P_J^\pm \mathcal{S}$. 

A Hilbert space $\mathcal{H}_J^\text{p}$ can then be constructed from the Cauchy completion of $P_J^{+}\mathcal{S}$ with respect to the norm induced by the inner product \eqref{innerprod}, which we will refer to as \emph{one-particle Hilbert space}. Also, from the properties of $J$, it is straightforward to check that $(P_J^{-}\mathcal{S})^*=P_J^{+}\mathcal{S}^*$, and hence that the completion of $(P_J^{-}\mathcal{S})^*$ with respect to the complex conjugate of \eqref{innerprod} is a Hilbert space, denoted by $\mathcal{H}_J^\text{ap}$ and called \emph{one-antiparticle Hilbert space}. The full Hilbert space of the theory is the direct sum of both spaces,
$\mathcal{H}_J=\mathcal{H}_J^\text p\oplus\mathcal{H}_J^\text{ap}$.

Let $\qty{\psi^\text p_n}_n$ be an orthonormal basis set of the one particle Hilbert space $\mathcal{H}_J^\text{p}$, and let  $\{\psi^\text{ap*}_n\}_n$ be a basis set of $\mathcal{H}_J^\text{ap}$. Hence $\qty{\psi^\text p_n,\psi^\text{ap}_n}_n$ is a basis set of the entire  space  of solutions. Any (complex) solution of the equations of motion can be written as
\begin{align}
    \phi(x)=\sum\limits_{n}\qty[b_n\psi^\text p_n(x)+d^*_n\psi^\text{ap}_n(x)],
    \label{modecomp}
\end{align}
    where the complex coefficients $b_n,d_n\in\mathbb{C}$ are given by
\begin{align}
    b_{n}=\innp{\psi^\text p_n}{P_J^{+}\phi},\quad   d_{n}^*=\innp{\psi^\text {ap}_n}{P_J^{-}\phi}.
    \label{sum}
    \end{align}
For definiteness, here we are assuming that the index $n$ runs over a discrete set of values.  In the case of a continuous basis set, the sums in \eqref{sum} and in the rest of this and the following sections can be regarded as integrals without affecting the discussion.
    
The coefficients $b_n$ are usually called ``annihilation-like'' variables for particles, and the coefficients $d_n^*$ are ``creation-like'' variables for antiparticles.
The Poisson bracket structure of the theory induces the following Poisson algebra for these coefficients,
\begin{align}
    \poissonbracket{b_n}{b^*_m}=\poissonbracket{d_n}{d^*_m}=-i\delta_{nm},\quad \poissonbracket{b_n}{d_m}=\poissonbracket{b_n}{d_m^*}=0.
    \label{bbddPBrels}
\end{align}

We can now proceed to quantize the field.
We identify the one-particle Fock space of particles and antiparticles with $\mathcal{H}_J=\mathcal{H}_J^\text p\oplus\mathcal{H}_J^\text{ap}$. The full Fock space of the quantum theory, $\mathfrak{F}(\mathcal{H}_J)$, is constructed straightforwardly from $\mathcal{H}_J$ by tensor products with the adequate symmetrization.
The classical creation and annihilation-like variables are then promoted  to creation and annihilation operators on Fock space, such that we obtain a representation of the canonical commutation relations associated to the classical Poisson brackets \eqref{bbddPBrels}:
\begin{align}
[\hat b_n,\hat b_m^\dagger]=\delta_{nm},\quad [ d_n,\hat d_m^\dagger]=\delta_{nm}.
 \label{creannh}
\end{align}
The Fock vacuum is defined as the state annihilated by all annihilation operators. On the other hand,
the quantum field operator $\hat\phi$ associated to the classical complex field is defined by
\begin{align}
    \hat \phi(x)=\sum_n[\hat b_{n}\psi_n^\text{p}(x)+\hat d^\dagger_{n}\psi_n^\text{ap}(x)],
    \label{quantfield}
    \end{align}
which is the quantum version of \eqref{modecomp}.

An important feature of this Fock space construction is its dependence on the choice of complex structure $J$. Note that every complex structure defines a particular splitting of the one-particle Hilbert space into two mutually orthogonal sectors, one for particles and one for antiparticles, each of them with their own sets of creation and annihilation operators. Therefore, this choice of $J$ ultimately defines the particular representation of the canonical commutation relations, determining uniquely the quantization of the classical theory. It is physically reasonable to require that the complex structure $J$   be compatible with the symmetries of the classical system, so that these are implemented unitar\-ily in the quantum theory.  In this way, the associated vacuum state is invariant under such unitary transformations. For example,
in the case of quantum field theory in flat spacetime, and in the absence of a background gauge field, 
the Poincaré symmetry group 
induces a unique preferred choice of complex structure, which preserves this symmetry group. 

However, in the presence of sufficiently strong and/or time-dependent external backgrounds, the Poincaré symmetry is generally not present, and a unitary implementation of the symmetries is not enough to remove the ambiguity in the choice of complex structure. As  a consequence, in these situations, an interpretation of the field states in terms of physical particles becomes difficult, unless extra conditions are imposed as criteria in order to choose a particular complex structure (or, equivalently, a particle-antiparticle decomposition). A physically reasonable condition to impose is the unitary implementability of time evolution in the  Fock space,  in order to have a  well-defined  particle interpretation at every time during the evolution of the system   up to unitary equivalence  \cite{Cortez:2015mja, Cortez_2020}.

\section{Canonical transformations}
\label{canonical}

In this section, we review how the ambiguity in the choice of a complex structure can be  related to the freedom of performing  canonical transformations in the classical theory. We also show how this fact translates, in the quantum theory, to the existence of different possibly inequivalent quantizations, and how imposing unitary implementation of both the symmetries and the evolution can reduce this ambiguity. Finally, we discuss the ambiguities due to gauge freedom in the description of the classical background.

\subsection{Bogoliubov transformations and unitary implementation}
\label{Bogoliubov}

The mode decomposition of $\phi(x)$ in \eqref{modecomp} strongly depends on the particular choice for the complex structure. Had we chosen a different complex structure $\tilde{J}$ with associated annihilation-like variables for particles $\tilde{b}_n$, and creation-like variables for antiparticles $\tilde{d}^*_n$, the procedure described in the last part of section \ref{classicalscalar} would have been the same, yielding a different decomposition for the field solutions:
\begin{align}
    \phi=\sum\limits_{n}[\tilde{b}_n\tilde{\psi}^\text{p}_n+\tilde{d}^*_n\tilde{\psi}^\text{ap}_n],
    \label{phi2}
\end{align}
where $\{\tilde{\psi}^\text{p}_n\}_n$ is an orthonormal basis set of the one particle Hilbert space $P_{\tilde{J}}^{+}\mathcal{S}$, and $\{\tilde{\psi}^\text{ap}_n\}_n$ is a basis set of $(P_{\tilde{J}}^{-}\mathcal{S})^*= P_{\tilde{J}}^{+}\mathcal{S}^*$. Since both $\qty{\psi^\text{p}_n,\psi^\text{ap}_n}_n$ and $\{\tilde{\psi}^\text{p}_n,\tilde{\psi}^\text{ap}_n\}_n$  provide bases for the same space of solutions, we can express the elements of one of the basis in terms of the elements of the other:
\begin{align}
    \psi^\text{p}_n=\sum_m[\alpha^\text{p}_{nm}\tilde{\psi}^\text{p}_m+\beta^\text{p}_{nm}\tilde{\psi}^\text{ap}_m],\quad \psi^\text{ap}_n=\sum_m[\beta^\text{ap}_{nm}\tilde{\psi}^\text{p}_m+\alpha^\text{ap}_{nm}\tilde{\psi}^\text{ap}_m].
    \label{basischange}
\end{align}
Note that the so-called $\beta$-coefficients mix the particle and antiparticle sectors, and therefore, as long as they are not all vanishing, the concepts of particles and antiparticles in the theory with complex structure $J$ is different from that of the theory defined by $\tilde{J}$.

Inserting \eqref{basischange} into \eqref{modecomp} and comparing with \eqref{phi2}, one finds that the corresponding creation and annihilation variables are related through the transformation
\begin{align}
    \tilde{b}_m=\sum_n[\alpha^\text{p}_{nm}b_n+\beta^\text{ap}_{nm}d^*_n],\quad \tilde{d}^*_m=\sum_n[\beta^\text{p}_{nm}b_n+\alpha^\text{ap}_{nm}d^*_n].
    \label{newcreanihilation}
\end{align}

The requirement that the annihilation and creation variables so defined must satisfy the same Poisson bracket relations in Eq. \eqref{bbddPBrels} imposes the following conditions on the transformation coefficients:
\begin{align}
    \sum_j[\alpha_{jn}^\text{p}\alpha_{jm}^{\text{p}*}-\beta_{jn}^\text{ap}\beta_{jm}^{\text{ap}*}]&=\delta_{nm}=\sum_j[\alpha_{jn}^\text{ap}\alpha_{jm}^{\text{ap}*}-\beta_{jn}^\text{p}\beta_{jm}^{\text{p}*}],
       \nonumber\\
        \sum_j[\alpha_{nj}^\text{p}\alpha_{mj}^{\text{p}*}-\beta_{nj}^\text{ap}\beta_{mj}^{\text{ap}*}]&=\delta_{nm}=\sum_j[\alpha_{nj}^\text{ap}\alpha_{mj}^{\text{ap}*}-\beta_{nj}^\text{p}\beta_{mj}^{\text{p}*}],
       \nonumber\\
    \sum_j[\alpha_{jn}^\text{p}\beta_{jm}^{\text{p}*}-\beta_{jn}^\text{ap}\alpha_{jm}^{\text{ap}*}]&=0=  \sum_j[\alpha_{nj}^\text{p}\beta_{mj}^{\text{p}*}-\beta_{nj}^\text{ap}\alpha_{mj}^{\text{ap}*}].
    \label{alphabetacond}
\end{align}
Hence, any transformation of the form \eqref{basischange} that satisfies  \eqref{alphabetacond} is a canonical transformation of the field variables, since it preserves the symplectic structure. Transformations of this kind are known as (classical) Bogoliubov transformations \cite{wald1994quantum}.

Let us study now how the Bogoliubov transformation that maps the set of variables $\{b_n,d^*_n\}_n$ to the set $\{\tilde{b}_n,\tilde{d}^*_n\}_n$ translate into the quantum theory.  Let us call $\hat{\tilde\phi}$ the field operator in the theory defined by $\tilde{J}$, namely resulting from promoting $\tilde{b}_n$ and $\tilde{d}_n^*$ in \eqref{phi2} to annihilation and creation operators. This Bogoliubov transformation is said to be unitarily implementable in the quantum theory if there exists a unitary operator $\hat{U}:\mathfrak{F}(\mathcal{H}_J)\rightarrow \mathfrak{F}(\mathcal{H}_J)$ such that $\hat{\tilde \phi}=\hat{U}\hat{\phi}\hat{U}^\dagger$,
in which case $\hat{\tilde\phi}$ belongs to $\mathfrak{F}(\mathcal{H}_J)$ or, in other words, $\mathfrak{F}(\mathcal{H}_{\tilde{J}})=\mathfrak{F}(\mathcal{H}_J)$. If such unitary operator  exists the two complex structures $J$ and $\tilde{J}$ are said to be unitarily equivalent, and define unitarily equivalent quantizations.

A way to characterize the unitary transformations that can be unitarily implemented is a theorem by Shale \cite{Shale:1962,Honegger:1996,RUIJSENAARSBogoII}, which states that the 
Bogoliubov transformation admits a unitary implementation in the space of Fock states if and only if
\begin{align}
    \sum_{n,m}\left(\abs{\beta^\text{p}_{nm}}^2+\abs{\beta^\text{ap}_{nm}}^2\right)<\infty,
    \label{squaresumcondition}
\end{align}
i.e. the $\beta$-coefficients must be square summable.

It turns out that not every canonical transformation can be implemented as a unitary operator \cite{RUIJSENAARSBogoII}. The fact that the Bogoliubov transformation that relates two given complex structures cannot be unitarily implemented gives rise to two unitarily nonequivalent quantizations of the classical field variables. This is the ambiguity that we have previously referred to. Thus, some conditions must be imposed on the set of complex structures in order to eliminate, or at least to reduce, this ambiguity.

So far, we have considered only those canonical transformations that relate two sets of field variables from the same space of solutions $\mathcal{S}$. 
Nevertheless, one can argue that the most general canonical transformation does not need to be an endomorphism in the solution space, since the two sets of canonical variables related through a general time-dependent canonical transformation will not satisfy the same equations of motion in general. Therefore, we must consider also canonical transformations that relate two different spaces of solutions. 

Being more specific, in the above discussion the sets $\{\tilde\psi_n^\text{p},\tilde\psi_n^\text{ap}\}_n$ could provide bases for $P^+_{\tilde J}\mathcal S'$ and $(P^-_{\tilde J}\mathcal S')^*$ respectively, with $\mathcal S'$ being another space of solutions, related to $\mathcal S$ by means of a time-dependent canonical transformation. Then, in Eq. \eqref{phi2}, the annihilation and creation-like coefficients would acquire a particular time-dependence, so that $\phi$ remains to be an element of $\mathcal{S}$. 

An example of these transformations in the context of quantum fields in cosmology, corresponds to different parametrizations of the field variables related by a rescaling in terms of the scale factor of the spacetime \cite{Cortez:2015mja}. These transformations  might be  actually needed in order to achieve a unitary implementation of the dynamics in the quantum theory, as we will discuss in the next section.

\subsection{Evolution as a Bogoliubov transformation}

 Let us now regard the evolution as a Bogoliubov transformation and further discuss the implications that the requirement of its unitary implementation may have for the complex structure. Here we closely follow the discussion in \cite{Cortez:2015mja}.

\begin{figure} 
    \centering
    \includegraphics[width=0.49\textwidth]{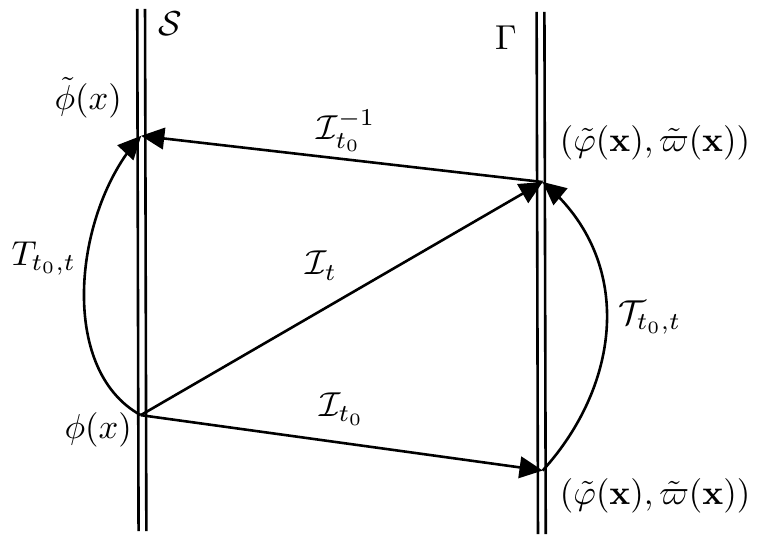}
    \hspace{0.05cm}
    \caption{Diagram representing the isomorphic relation between the covariant phase space $\mathcal{S}$ and the canonical phase space $\Gamma$. The evolution operator $\mathcal{T}_{t_0,t}$ relates two elements of $\Gamma$, which can be understood as the initial conditions for a certain solution $\phi\in\mathcal{S}$ given at two different times.}
    \label{fig:symplecisom}
\end{figure}

 As previously mentioned, for any time $t$ there is a symplectic isomorphism $\mathcal{I}_t$ between the covariant phase space $\mathcal{S}$ and the canonical phase space $\Gamma$, such that 
$
         \mathcal{I}_t\phi= (\varphi(\vec{x}),\varpi(\vec{x})) 
$, where $\varphi(\vec{x})=\phi(x)|_{\Sigma_t}$ is the field evaluated at time $t$, and $\varpi(\vec{x})=\pi(x)|_{\Sigma_t}$ is the canonically conjugate momentum of the field, also at time $t$, instant that defines the Cauchy hypersurface $\Sigma_t$. We can once and for all fix an arbitrary initial time $t_0$ to identify $\mathcal{S}$ with $\Gamma$. Then the classical evolution operator on the Cauchy space from the initial time $t_0$ to an arbitrary time $t$ is given by $\mathcal{T}_{t_0,t}\equiv \mathcal{I}_{t} \mathcal{I}_{t_0}^{-1}$, as illustrated in figure \ref{fig:symplecisom}.
In the covariant phase space $\mathcal{S}$, the evolution is given by the transformation 
\begin{align}
    \phi\mapsto\tilde\phi\equiv \mathcal{I}^{-1}_{t_0}\mathcal{T}_{t_0,t}\mathcal{I}_{t_0}\phi=\mathcal{I}^{-1}_{t_0}\mathcal{I}_{t}\phi\equiv T(t_0,t)\phi.
    \label{evolcanonical}
\end{align}
Since both $\mathcal{I}_{t}$ and $\mathcal{I}_{t_0}$ are symplectic isomorphisms, the map \eqref{evolcanonical} is indeed
a symplectomorphism, and therefore can be written as a (time-dependent) Bogoliubov transformation. 

Let us move now to the quantum theory, for which we introduce a complex structure $J$ on the covariant phase space $\mathcal{S}$.
The isomorphism $\mathcal{I}_t$ induces a one-parameter family of complex structures $\mathcal{J}_t$ on the canonical space $\Gamma$ given by $\mathcal{J}_t=\mathcal{I}_t J\mathcal{I}_t^{-1}$. In particular, at initial time $t_0$ we define $\mathcal{J}_{t_0}$.
Then the complex structure $\mathcal{J}_t$ generated by time evolution of $\mathcal{J}_{t_0}$ satisfies by construction $ \mathcal{J}_t= \mathcal{T}_{t_0,t} \mathcal{J}_{t_0} \mathcal{T}_{t_0,t}^{-1}$.
In turn, the family of complex structures $ \mathcal{J}_t:\Gamma\rightarrow\Gamma$ provides a one-parameter family of complex structures $J_t= \mathcal{I}_{t_0}^{-1} \mathcal{J}_t \mathcal{I}_{t_0}=T(t_0,t) J T^{-1}(t_0,t)$ on the covariant space $\mathcal{S}$. 
A unitary implementation of the evolution map \eqref{evolcanonical} then amounts to the condition that  $J=J_{t_0}$ and $J_t$ define unitarily equivalent quantum theories for all $t$. This in turn, will require the square summability of the $\beta$-coefficients corresponding to the evolution map \eqref{evolcanonical}.

Actually, it might happen that there is no complex structure on the covariant phase space $\mathcal{S}$ that allows for a unitary implementation of the dynamics. Nevertheless, as discussed at the end of section \ref{Bogoliubov}, we always have the freedom to consider a time-dependent canonical transformation mapping the space $\mathcal S$ to another space $\mathcal S'$, so that unitarity of the dynamics can be achieved in the Fock space constructed out of a convenient complex structure defined in $\mathcal S'$. This is precisely what previous works in cosmology pushed forward as criterion to remove the ambiguity in the choice of complex structure \cite{Cortez:2015mja}, and that we will later exploit in section \ref{system} when quantizing a charged scalar field coupled to a classical electric field. Finally, let us note that one can always transform the original field description, i.e. the canonical space $\mathcal S$, to a space $\mathcal{S'}$ with trivial dynamics, in which case the unitary implementation of the dynamics trivializes. We will avoid that possibility since unitarity of the dynamics does not longer serve as selection criterion (the identity is always unitary).

\subsection{Gauge transformations and physical equivalence}
\label{gauge}

Another example of time-dependent canonical transformations which is relevant to the case that we are studying are gauge transformations $g$ that belong to a gauge group $\mathcal G$ (in our case $U(1)$), and that map the potential $A_\mu$ to $A_\mu^g$.
Since in our framework we will not consider the external gauge field as dynamical, gauge transformations cannot be considered as symmetries (automorphisms of the covariant phase space). Instead, they can be treated as a set of canonical transformations between field variables corresponding to different covariant phase spaces, $\mathcal{S}_A$ and $\mathcal{S}_{A^g}$ respectively. Then, any gauge transformation can be described as an operator $ G(g):\mathcal{H}\rightarrow\mathcal{H}^g$, with $\mathcal{H}$ and $\mathcal{H}^g$ the one-particle Hilbert spaces associated to the theories with potential $A_\mu$ and $A_\mu^g$ respectively. 
The quantizations in the two different gauges will be in general different, but must be \emph{physically equivalent}. Namely, the values of all transition amplitudes must be the same in both theories. 
This is so if the operator $G(g)$ is unitary, namely if it verifies
\begin{align}
   \innp{ G(g)\phi}{ G(g)\psi}_{\mathcal{H}^g}= \innp{\phi}{\psi}_{\mathcal{H}},\quad \forall \phi,\psi \in\mathcal{H}.
   \label{condinduced}
\end{align}
The condition of physical equivalence is then equivalent to unitary equivalence between the one-particle Hilbert spaces $\mathcal H$ and $\mathcal H^g$.

One can check that, given a complex structure $J$ on $\mathcal{S}_A$, the corresponding complex structure induced by the gauge transformation $G(g)$ on $\mathcal{S}_{A^g}$, $J^g={G}(g)J{G}(g^{-1})$, render the two quantum theories physically equivalent. 

Indeed, consider the Hilbert space $\mathcal{H}_J^\text{p}$ as the completion of $P^+_{J}\mathcal{S}_A$ in the inner product 
\begin{align}
   \langle\psi,\phi\rangle_{\mathcal{H}_J^\text{p}}=\mu(\psi,J\phi), \quad\forall\phi,\psi\in\mathcal{H}_J^\text{p}, 
\end{align}
and $\mathcal{H}^{\text{p},g}_{J}$ as the completion of $P^+_{J^g}\mathcal{S}_{A^g}$ in the inner product 
\begin{align}
   \langle G(g)\psi,G(g)\phi\rangle_{\mathcal{H}^{\text{p},g}_{J}}=\mu(G(g)\psi,J^g G(g)\phi), 
\end{align}
then, using the definition for the induced  $J^g$  and the fact that $G(g)$ is unitary,

\begin{align}
        \langle{G}(g)\psi,{G}(g)\phi\rangle_{\mathcal{H}^{\text{p},g}_{J}}
       =\mu({G}(g)\psi,{G}(g)J\phi)=\mu(\psi,J\phi)=\langle\psi,\phi\rangle_{\mathcal{H}^\text{p}_J},
\end{align}
as desired for the one-particle Hilbert space of particles. To obtain this result we have used the fact that the sesquilinear form $\mu$ is gauge independent. One would proceed similarly with the one-particle Hilbert space of antiparticles, arriving at a similar conclusion.

Note that the previous construction implies that gauge transformations are implemented trivially in the Fock space of the theory, namely $\hat{G}(g):\mathfrak{F}(\mathcal{H}_J)\rightarrow\mathfrak{F}(\mathcal{H}_{J}^g)$, or in other words $\hat\phi^g=G(g)\hat\phi$,
preserving the gauge equivalence principle in the quan\-tum theory.

\section{Scalar field in a homogeneous, time dependent electric field}
\label{system}

After the previous considerations, we are now ready to consider the Fock quantization of the system we are interested in: a charged scalar field in a flat spacetime with a spatially homogeneous time-dependent electric field background. We will study the unitary implementability of the resulting quantum dynamics.

Eq. \eqref{ChargedKG} is the equation of motion for a massive complex scalar field coupled to an external electromagnetic background, for an arbitrary potential $A_\mu(x)$. 
In the particular case of a spatially homogeneous ---possibly time-dependent--- external electric field, this equation of motion reduces to
\begin{align}
   \qty{-\partial^2_t+\vec{\nabla}^2-m^2+2iq\vec{A}(t)\cdot\vec{\nabla}-q^2 {A}(t)^2} \phi(\vec{x},t)=0,
   \label{eqderivs}
\end{align}
where $A=|\vec A|$.
Note that we have already fixed the gauge to the so-called temporal gauge, in which  $A_\mu(x)=(0,\Vec{A}(t))$. There is still a residual gauge freedom that we will fix by imposing $\Vec{A}(t_0)=0$, so that in the absence of electromagnetic field this equation reduces to the Klein-Gordon equation.
In this section, we will study the unitary implementation of the evolution in this temporal gauge, and, as we have seen in the previous section, this will induce a unique  procedure to implement the unitary evolution in any other gauge.

Due to the independence of the potential vector on the spatial coordinates, spatial translations are still a symmetry of the equations of motion, so we may expand the solutions of \eqref{eqderivs} in the basis of plane wave modes:
\begin{align}
   \phi(\vec{x},t)=\int_{\mathbb{R}^3}d\vk^3 \, e^{i\vec{k}\cdot\vec{x}}\varphi_{\vec{k}}(t),
    \label{modexpansions}
\end{align}
This expansion significantly simplifies  the description of the field dynamics, since each (complex) mode function $\varphi_\vk(t)$, labeled by its wave vector $\vk\in\mathbb{R}^3$, becomes decoupled from the others. 
Furthermore,  the discussion of the dynamics is further simplified by splitting the field modes into their real and imaginary parts, 
\begin{align}
    \varphi_\vk(t)=\tfrac{1}{\sqrt{2}}\qty[\xi_\vk(t)+i\eta_\vk(t)],
\end{align}
because both of them  satisfy the same mode-dependent equation of motion:
\begin{align}
    \Ddot{z}_\vk+(k^2+m^2+2qkA_k+q^2 A^2)z_\vk=0,
    \label{modeqmotion}
\end{align}
where $z_\vk=(\xi_\vk,\eta_\vk)$, $k=\abs{\vk}$,  $A=\abs{\vA}$, and $A_k=\vA\cdot \vu{k}$ is the component of $\vA$ along the direction $\vu{k}=\vk/k$.
The dynamics of the complex field can therefore be written in terms of a pair of real variables for each mode that evolve independently but according to the same equation of motion \eqref{modeqmotion}.

We will now study the classical dynamics of these real mode functions, in order to understand their asymptotic behavior in the ultraviolet regime of large wave number $k$. Our goal is to find the conditions that a complex structure must satisfy for the classical evolution to be implemented as a unitary transformation in the quantum theory.

\subsection{Classical evolution in terms of initial conditions}
\label{sectA}
For a start, let us note that 
the unique nontrivial equal-time canonical Poisson bracket between the field modes and their derivatives is
$
   \{\varphi_\vk(t),\dot{\varphi}_{\vk'}^*(t)\}=\delta_{\vk\vk'}
$.
Thus, the real and the imaginary parts of the field modes $z_\vk$ satisfy
$
    \{z_\vk,\dot{z}_{\vk'}\}=\delta_{\vk\vk'}
$. This means that the canonical phase space of the theory is then given by 
$
    \Gamma=\qty{(z_\vk(t_0), \dot{z}_\vk(t_0))} 
$ 
for some initial time $t_0$.
From these initial conditions, 
the values of   $(z_\vk(t),\dot{z}_\vk(t))$  at any time $t$ is obtained via
\begin{align}
    \mqty(z_\vk(t)\\ \dot{z}_\vk(t))={T}_\vk(t_0,t)\mqty(z_\vk(t_0)\\ \dot{z}_\vk(t_0)),
    \label{evolutionmodes}
\end{align}
where the evolution operator  ${T}_\vk(t_0,t)$ is a $2\times 2$ matrix of operators $\mathcal{U}$ and $\mathcal{V}$ to be determined from Eq. \eqref{modeqmotion}:
\begin{align}
    {T}_\vk(t_0,t)=\mqty(\mathcal{U}_\vk^1(t_0,t)&\mathcal{U}_\vk^2(t_0,t)\\\mathcal{V}_\vk^1(t_0,t)&\mathcal{V}_\vk^2(t_0,t)).
    \label{eq:TUVmatrix}
\end{align}

Let   $\zeta_\vk=e^{i\Theta_\vk(t)}$ be a complex solution of \eqref{modeqmotion}, with $\Theta_\vk$  a sufficiently smooth complex function. Then any real solution of that equation (and its time derivative) can be written as
\begin{align}
    &z_\vk(t)=C_\vk e^{i\Theta_\vk(t)}+C_\vk^*e^{-i\Theta^*_\vk(t)},\nonumber\\ &\dot{z}_\vk(t)=i\qty[C_\vk\dot{\Theta}_\vk(t)e^{i\Theta_\vk(t)}-C_\vk^*\dot{\Theta}^*_\vk(t) e^{-i\Theta^*_\vk(t)}],
    \label{generalsolutions}
\end{align}
where $C_\vk$ is a complex constant.
These relations allow us to write $C_\vk$  in terms of the initial conditions $z_\vk(t_0)=z_{\vk,0},\,\Theta_\vk(t_0)=\Theta_{\vk,0}$, and their time derivatives, which in turn lead  to the following expressions for the functions $z_\vk$ and $\dot{z}_\vk$,
\begin{align}
    &z_\vk(t)=2\Re\bigg[\frac{(\dot{\Theta}^*_{\vk,0}z_{\vk,0}-i\dot{z}_{\vk,0})}{\dot{\Theta}^*_{\vk,0}+\dot{\Theta}_{\vk,0}}e^{i(\Theta_\vk(t)-\Theta_{\vk,0})}\bigg],\nonumber\\
    &\dot{z}_\vk(t)=2\Re\bigg[\frac{(i\dot{\Theta}^*_{\vk,0}z_{\vk,0}+\dot{z}_{\vk,0})}{\dot{\Theta}^*_{\vk,0}+\dot{\Theta}_{\vk,0}}\dot{\Theta}_\vk(t)e^{i(\Theta_\vk(t)-\Theta_{\vk,0})}\bigg].
\end{align}

From these expressions one can easily read the matrix elements of the evolution operator ${T}_\vk(t_0,t)$ given in Eq.~\eqref{eq:TUVmatrix}:
\begin{align}
        &\mathcal{U}_\vk^1(t_0,t)=
        \frac{\Re\qty[\dot{\Theta}^*_{\vk,0}e^{i(\Theta_\vk(t)-\Theta_{\vk,0})}]}{\Re(\dot{\Theta}_{\vk,0})},\quad \mathcal{V}_\vk^1(t_0,t)=-\frac{\Im\qty[\dot{\Theta}^*_{\vk,0}\dot{\Theta}_\vk(t)e^{i(\Theta_\vk(t)-\Theta_{\vk,0})}]}{\Re({\dot{\Theta}_{\vk,0}})}, \nonumber\\ &\mathcal{U}_\vk^2(t_0,t)=\frac{\Im\qty[e^{i(\Theta_\vk(t)-\Theta_{\vk,0})}]}{\Re({\dot{\Theta}_{\vk,0}})},\quad\qquad \mathcal{V}_\vk^2(t_0,t)=\frac{\Re\qty[\dot{\Theta}_{\vk}(t)e^{i(\Theta_\vk(t)-\Theta_{\vk,0})}]}{\Re({\dot{\Theta}_{\vk,0}})}.
    \label{evolmatrixcomponents}
\end{align}

\subsection{Asymptotic analysis of the solutions in the ultraviolet regime}
\label{asymp}

In order to study the behavior of the mode functions at large values of the wave number  $k$, we do not need the explicit expression of the general solution. Instead, it is enough to realize (as shown below) that the solution for $\Theta_\vk(t)$ can be written as
\begin{align}
     \Theta_\vk(t)=\theta_\vk(t)+\int_{t_0}^t\Lambda_\vk(\tau)d\tau, \quad \theta_\vk(t)=-\int_{t_0}^t d\tau\omega(\tau),
     \label{theta}
\end{align}
where  $\Lambda_\vk(t)$ is a function of $t$ of asymptotic order   $\mathcal{O}(k^{-1})$,  $\forall t\in\mathbb{R}$, with initial value $\Lambda_\vk(t_0)=0$, and
\begin{align}
  \omega(t)=\sqrt{(\vk+q\vA)^2+m^2}.\label{omegat}
\end{align}
Note that here we have chosen for convenience and without loss of generality the initial condition $\Theta_\vk(t_0)=0$.
A different initial condition would simply introduce an irrelevant constant phase in the mode solutions.

This  implies that $\dot{\Theta}_\vk(t_0)=-\sqrt{k^2+m^2}=-\omega_0 $  (since we have chosen $\vA(t_0)=0$)
and the components of the evolution matrix simplify to:
\begin{align}
        \mathcal{U}_\vk^1(t_0,t)&=\Re\qty[e^{i\Theta_\vk(t)}],\,\,\quad\,\,\, \mathcal{V}_\vk^1(t_0,t)=-\Im\qty[\dot{\Theta}_\vk(t)e^{i\Theta_\vk(t)}],\nonumber\\
        \mathcal{U}_\vk^2(t_0,t)&=-\frac{\Im\qty[e^{i\Theta_\vk(t)}]}{\omega_0},\quad
        \mathcal{V}_\vk^2(t_0,t)=-\frac{\Re\qty[\dot{\Theta}_{\vk}(t)e^{i\Theta_\vk(t)}]}{\omega_0}.
\end{align}

In the rest of this section we will indeed show that \eqref{theta} gives the correct behavior for $\Theta_\vk$. Inserting $z_\vk=\exp(i\Theta_\vk)$ into \eqref{modeqmotion}, and considering \eqref{theta}, we find a Riccati equation  for the function $\Lambda_\vk$:
\begin{align}
    \dot{\Lambda}_\vk=-i[\Lambda^2_\vk-2\omega(t)\Lambda_\vk]+\dot\omega(t).
\end{align}
Note that $\dot \omega$ is $\mathcal{O}(k^0)$.
By assumption, we have $\Lambda_\vk=\mathcal{O}(k^{-1})$ in the ultraviolet regime (the self-consistency of this behavior will be checked later). Then we can neglect the term proportional to $\Lambda_\vk^2$ with respect to the one proportional to $\omega\Lambda_\vk$, and hence $\Lambda_\vk$ must asymptotically behave (in the ultraviolet regime) as the function $\tilde{\Lambda}_\vk(t)$ which satisfies
\begin{align}
    \dot{\tilde{\Lambda}}_\vk=2i\omega\tilde{\Lambda}_\vk+\dot\omega(t).
\end{align}
For convenience, we choose the same initial condition for $\tilde\Lambda_\vk$ as for $\Lambda_\vk$, namely $\tilde\Lambda_\vk(t_0)=0$. Then $\tilde\Lambda_\vk$ has the form
\begin{align}
    \tilde{\Lambda}_\vk(t)=e^{-2i\theta_\vk(t)} \int_{t_0}^t d\tau \dot\omega(\tau)e^{2i\theta_\vk(\tau)},
    \label{intparts}
\end{align}
with $\theta_\vk(t)$ defined in \eqref{theta}. Note that both $\theta_\vk(t)$ and $\dot{\theta}_\vk(t)=-\omega(t)$ are $\mathcal{O}(k)$ for all $t$.

Integration by parts of equation \eqref{intparts} yields:
\begin{align}
        \tilde{\Lambda}_\vk(t)
        =& \frac{i}{2}\Bigg\{\frac{ \dot\omega(t)}{\omega(t)}
        -e^{-2i\theta_\vk(t) }\bigg[\frac{ \dot\omega(t_0)}{\omega_0} 
         +\int_{t_0}^t d\tau e^ {2i\theta_\vk(\tau) } \dv{}{\tau}\frac{ \dot\omega(\tau)}{\omega(\tau)}\bigg] \Bigg\},
    \label{ultrvreg}
\end{align}
and given that  $\dot\omega/\omega=\mathcal{O}(k^{-1})$, from \eqref{ultrvreg} it is straightforward to see that $\abs*{\tilde{\Lambda}_\vk}$ behaves as $\mathcal{O}(k^{-1})$ in the ultraviolet limit ---and hence so does $\abs{\Lambda_\vk}$ as assumed---. In other words, there exists a positive, $k$-independent function $F(t)$ such that
\begin{align}
    \abs{\Lambda_\vk(t)}\leq\frac{F(t)}{k}
\label{req}
\end{align}
for any finite time $t$, if the integral in \eqref{ultrvreg} is well defined. In particular, this happens whenever  
$\dot\omega/\omega$ is finite.
Note that this condition imposes some mild restrictions on the set of possible electromagnetic potentials which can be considered for the hypothesis made in \eqref{theta} to hold.

\subsection{Fock quantization and unitary implementation of the dynamics}
As we have seen in the previous sections, there is no unique choice for the complex structure which implements the definition of creation/annihilation variables in the classical theory. After proceeding with the canonical quantization of the field, this translates into the non uniqueness of a privileged vacuum state. This ambiguity is particularly problematic whenever the choice of two different complex structures results in  two vacuum states that are not unitarily equivalent, because it implies that both complex structures define two different quantizations that are not directly comparable.

On the other hand, a fundamental property of time evolution in quantum mechanics is that it must be implemented in the Hilbert space (in Schrödinger's picture) or the space of operators (in Heisenberg's picture) by means of a unitary operator. With this in mind, and in order to reduce the ambiguity in the choice of complex structures, we will consider as physically relevant only those which satisfy the following natural conditions:
\begin{enumerate}
    \item The complex structure preserves the symmetries of the equations of motion.
    \item The complex structure allows the unitary implementation of the time evolution.
\end{enumerate}
In the case under study, condition 1 implies that the complex structure (or, equivalently, the set of creation and annihilation variables which it defines) does not mix the dynamically decoupled modes $(\xi_\vk, \eta_\vk)$. Hence, the considered complex structures will be characterized by annihilation and creation-like variables of the form:
\begin{align}
    \mqty(a_\vk(t)\\a_\vk^*(t))=\mathfrak{J}_\vk^a(t)   \mqty(\xi_\vk(t)\\\dot{\xi}_\vk(t)),\quad \mqty(b_\vk(t)\\b_\vk^*(t))=\mathfrak{J}_\vk^b(t)   \mqty(\eta_\vk(t)\\\dot{\eta}_\vk(t)),
    \label{creationanhiCS(scalar)}
\end{align}
where
\begin{align}
    \mathfrak{J}_\vk^i(t)=\mqty(f^i_\vk(t)&g^i_\vk(t)\\f^{i*}_\vk(t)&g_\vk^{i*}(t)),\quad i=a,b.
    \label{Jcomplex}
\end{align}
The matrix operators $\mathfrak{J}_\vk^i(t)$ are uniquely defined by the chosen complex structure.
At the classical level, equation \eqref{creationanhiCS(scalar)} describes a canonical transformation from the field modes (and their canonically conjugate momenta) to the corresponding creation/annihilation variables. 
The full transformation is a block diagonal matrix, and the real and imaginary parts of the field remain decoupled at every time.
Note that, in the most general case, this transformation is mode-dependent, as well as time-dependent.

Since the pairs $(a_\vk(t),a_\vk^*(t))$ and $(b_\vk(t),b_\vk^*(t))$ represent two equivalent copies of the creation/annihilation varables of a scalar field, we will study only the unitary implementation of the evolution for one of them ---namely, $(a_\vk(t),a_\vk^*(t))$---, and the result applies in a  straightforward manner to the other. To simplify the notation, we will omit the superindex $i$ in $ \mathfrak{J}_\vk^i(t).$

For the variables $(a_\vk(t),a_\vk^*(t))$ to preserve the canonical Poisson algebra relations $\{a_\vk,a^*_\vk\}=-i$ at every time, the components $f_\vk,g_\vk$ of the transformation matrix $\mathfrak{J}_\vk(t)$ must satisfy
\begin{align}
    f_\vk(t) g_\vk^*(t)-g_\vk(t) f_\vk^*(t)=-i\quad \forall t.
    \label{conditionCS}
\end{align}
Also, condition 2 above imposes that the time evolution must be implemented as an unitary Bogoliubov transformation of the form
\begin{align}
    \mqty(a_\vk(t)\\a_\vk^*(t))=\mathfrak{B}_\vk(t_0,t)   \mqty(a_\vk(t_0)\\a_\vk^*(t_0)),
    \end{align}
   where 
   \begin{align}
         \quad\mathfrak{B}_\vk(t_0,t)=\mqty(\alpha_\vk(t_0,t)&\beta_\vk(t_0,t)\\\beta^*_\vk(t_0,t)&\alpha^*_\vk(t_0,t)). 
   \end{align}

From \eqref{creationanhiCS(scalar)} and \eqref{evolutionmodes}, we can obtain an explicit expression for $\mathfrak{B}_\vk(t_0,t)$:
\begin{align}
    \mathfrak{B}_\vk(t_0,t)=\mathfrak{J}_\vk(t)\mathcal{T}_\vk(t_0,t)\mathfrak{J}_\vk^{-1}(t_0),
\end{align}
which in turn allows us to read  the particular expression for the $\beta_\vk$ coefficients:
 \begin{align}
      \beta_\vk(t_0,t)=\frac{1}{2\omega_0}
\Big\{&[f_\vk(t)-ig_\vk(t)\dot{\Theta}^*_\vk(t)][f_\vk(t_0)-i\omega_0g_\vk(t_0)]e^{-i\Theta^*_\vk(t)}\nonumber\\
      &-[f_\vk(t)+ig_\vk(t)\dot{\Theta}_\vk(t)][f_\vk(t_0)+i\omega_0g_\vk(t_0)]e^{i\Theta_\vk(t)}\Big\}.
      \label{betacoefs}
\end{align}

As we have previously seen, the unitary implementability of a Bogoliubov transformation implies directly the square summability of the $\beta$-coefficients \eqref{squaresumcondition}. In this case, the (vector) index $\vk$ plays the role of the subindices in \eqref{squaresumcondition},  and since the components of $\vk$ take continuous values, one should understand the sum as the integral $\int_{\mathbb{R}^3}d\vk$. Thus, in order to analyze the square integrability of our $\beta$-coefficients, we need to study the ultraviolet behavior of $\abs{\beta_\vk}$. 
Taking into account that $\dot{\Theta}_\vk(t)=-\omega(t)+\mathcal{O}(k^{-1})$, we can write the modulus of the $\beta$-coefficients as
\begin{align}
\abs{\beta_\vk(t_0,t)}=\frac{1}{2\omega_0}\Big\lvert&\mathcal{F}^+_\vk(t) \mathcal{F}^-_\vk(t_0)e^{-i\theta_\vk(t)}
-\mathcal{F}^-_\vk(t)\mathcal{F}^+_\vk(t_0)e^{i\theta_\vk(t)}\nonumber\\&
+2i g_\vk(t)g_\vk(t_0)\sin[\theta_\vk(t)]\mathcal{O}(k^{0})\Big\rvert\left[1+\mathcal{O}(k^{-2})\right],
\label{assymptbeta}
\end{align}
with $\mathcal{F}^\pm_\vk(t)=f_\vk(t)\pm ig_\vk(t)\omega(t)$.
In the last line of \eqref{assymptbeta} we have used that $f_\vk\cdot g_\vk=\mathcal{O}(k^0)$, as deduced from \eqref{conditionCS}. Note that, as we have not justified yet the asymptotic order required for $g_\vk(t) g_\vk(t_0)$, we keep this contribution for the time being.

The condition of square integrability for these coefficients for each $t$  requires that their norm should be at least $\mathcal{O}(k^{-2})$ in order to compensate for the degeneracy (an integral in $\vk$ of terms that only depend on $k$ grows as $k^2$). This behavior
then translates into a set of restrictions for the functions $f_\vk(t)$ and $g_\vk(t)$ that characterize the complex structure. 
Indeed, the dominant terms must vanish identically in the asymptotic limit  $k\rightarrow \infty$. Moreover, the dominant terms coming from the first and second terms of \eqref{assymptbeta} cannot compensate one with another, since this  would be possible only by trivializing the quantum dynamics (possibility that we disregard). Therefore, they must vanish independently. These considerations, together with the normalization condition \eqref{conditionCS}, make us conclude that for each $t$ we need to require
\begin{align}
    \lim_{k\rightarrow\infty} \mathcal{F}^+_\vk(t)=0.
\end{align}
This motivates that $f_\vk(t)$ and $g_\vk(t)$ are related via
\begin{align}
    f_\vk(t)= - i\omega(t) g_\vk(t) [h_\vk(t)]^{-2},
    \label{conditionsuma}
\end{align}
with $h_\vk(t)$ a function that verifies
\begin{align}
    \lim_{k\rightarrow\infty} h_\vk(t)=1.
\end{align}
Now, if we substitute the behavior for $f_\vk(t)$ given in \eqref{conditionsuma} into \eqref{conditionCS}, we obtain
\begin{align}
   2\omega(t)\abs{g_\vk(t)}^2\text{Re}\{[h_\vk(t)]^{-2}\}=1,
\end{align}
which must be valid for all $k$,
This condition together with \eqref{conditionCS} fixes completely (up to a phase) the $f_\vk$ and $g_\vk$ functions. 
For simplicity, and without loss of generality, we may choose $g_\vk(t)$ and $h_\vk(t)$ to be real functions. Then, we find
\begin{align}
    g_\vk(t)=\frac{1}{\sqrt{2\omega(t)}}h_\vk(t),\quad  f_\vk(t)=-i\sqrt{\frac{\omega(t)}{2}}[h_\vk(t)]^{-1},
    \label{generalform}
\end{align}
which satisfy both \eqref{conditionCS} and \eqref{conditionsuma} by construction.
 Replacing these expressions back in \eqref{assymptbeta}, we obtain
\begin{align}
      \beta_\vk(t_0,t)=\frac{1}{4}
\Bigg\{&\left[h_\vk(t)-\frac{1}{h_\vk(t)}\right]\left[h_\vk(t_0)+\frac{1}{h_\vk(t_0)}\right]e^{-i\theta_\vk(t)}\\
      &-\left[h_\vk(t)+\frac{1}{h_\vk(t)}\right]\left[h_\vk(t_0)-\frac{1}{h_\vk(t_0)}\right]e^{i\theta_\vk(t)}\Bigg\}[1+\mathcal{O}(k^{-1})]+\mathcal{O}(k^{-2}).\nonumber
\end{align}
From this we see 
that the next to leading order of the function $h_\vk(t)$ should be $\mathcal{O}(k^{-2})$ to guarantee $\abs{\beta_\vk(t,t_0)}=\mathcal{O}(k^{-2})$.

In summary, up to a phase, the whole freedom in the choice of complex structure is encoded in the choice of the function $h_\vk(t)$, and the dynamics will be unitarily implementable as long as 
\begin{align}
     h_\vk(t)=1+\mathcal{O}(k^{-2})
     \label{h-cond}
\end{align}
is verified. A possible (and simplest) choice is
\begin{align}
   h_\vk(t)=1.
   \label{h}
\end{align}
In this case, the functions that characterize the complex structure acquire the simple form:
\begin{align}
    g_\vk(t)=\frac{1}{\sqrt{2\omega(t)}},\quad  f_\vk(t)=-i\sqrt{\frac{\omega(t)}{2}}.
    \label{simpleform}
\end{align}

We have thus found a particular complex structure, specified by Eq. \eqref{simpleform}, for which the evolution operator can be unitarily implemented in the Fock space.  Furthermore, our choice  of quantization agrees with that of previous approaches based on adiabatic vacua \cite{Kluger:1998bm,Fedotov2011, Huet2014}. Namely, from our perspective of unitarity of the quantum dynamics, adiabatic vacua are natural because they define a quantization in which the evolution is implemented unitarily in Fock space. Moreover, note that the above expression reduces to the usual complex structure for quantum fields  in flat spacetime when the potential vector goes to zero, so that the usual quantization for the field modes is recovered in the limit of vanishing external field. However, it is important to note that, whenever the potential contributes, we conclude that no complex structure independent of it will admit a unitary implementation of the dynamics. In particular, the complex structure usually chosen in flat spacetime and associated to the Minkowski vacuum corresponds to replacing $\omega(t)$ by $\omega_0$ in \eqref{simpleform}. This is obtained by choosing 
\begin{align}
     h_\vk(t)=\sqrt{\frac{\omega(t)}{\omega_0}}=1+\frac{qA_k(t)}{4k}+\mathcal{O}(k^{-2}),
\end{align}
which does not verify \eqref{h-cond}. Therefore, choosing a quantization based on the Minkowski vacuum does not allow to implement the dynamics unitarily in the case of a non-vanishing potential, in agreement with \cite{Ruijsencharged}.

Taking into account equations \eqref{simpleform} and recalling that  we have parametrized the complex solutions of the equations of motion as $\zeta_\vk=e^{i\Theta_\vk(t)}$, with our choice of complex structure we may rewrite the beta coefficients \eqref{betacoefs} as
\begin{align}
    \beta_\vk(t)=\frac{-1}{\sqrt{2\omega_0}}\qty{\sqrt{\frac{\omega(t)}{2}}\zeta_\vk^*(t)+i\frac{1}{\sqrt{2\omega(t)}}\dot{\zeta}_\vk^*(t)}.
\end{align}
Hence the squared modulus of these is given by
\begin{align}
    \abs{\beta_\vk}^2=\frac{1}{4\omega_0}\qty{\omega\abs{\zeta_\vk}^2+\omega^{-1}\abs*{\dot{\zeta}_\vk}^2+2\Im(\dot{\zeta}_\vk\zeta^*_\vk)},
    \label{betasquare}
\end{align}
and it is square integrable as we have shown (provided that the mild conditions in the potential that we have discussed at the end of section \ref{asymp} hold). Note that this expression for the beta coefficients, closely related to the particle number density, is similar to those obtained previously in the same context by means of the kinetic equation approach \cite{Fedotov2011,Huet2014}.

As a final remark, it is important to note that the choice of the particular function $h_{\vk}$ is still arbitrary (as long as it satisfies \eqref{h-cond}), and so will be the complex structure that it defines. As a consequence, there still remains some ambiguity in the quantization, which translates into an ambiguity in the particle number operator. This ambiguity cannot be fixed with the criterion of unitary dynamics alone, and more criteria are needed in order to completely remove the ambiguity on the particle number density for quantum fields in time-dependent backgrounds (possibly coming from experimental conditions). However, as we will show in the following section, the unitarity criterion allows to drastically reduce the initial ambiguity up to a unitary equivalence between quantizations ---which is not generally guaranteed, due to the absence of an analogue of the Stone-Von Neumann theorem in quantum field theory, as we have stated in previous sections---. 

\subsection{Uniqueness of the quantization}

In this section, we will show that the Fock representations defined by all of the complex structures that admit a unitary implementation of the evolution are unitarily equivalent. In other words, the criterion of unitary implementation of the dynamics indeed reduces the ambiguity of the quantization by selecting a unique Fock representation, up to unitary equivalence, similarly as what happens in cosmology \cite{Cortez:2016xsn,Cortez_2020,Cortez:2016oxm,Cortez:2016kwy,Cortez:2015mja, Cortez:2012cf,Gomar:2012xn}.

Let us consider a reference Fock representation $(a_\vk,a_\vk^*)$, defined by the map $\mathfrak{J}_\vk(t)$, and any other Fock representation $(\tilde{a}_\vk,\tilde{a}_\vk^*)$ defined by a different map $\tilde{\mathfrak{J}}_\vk(t)$, both satisfying the criteria of the previous section.
Both sets of annihilation and creation like variables are related, at any given time $t$, by the following Bogoliubov transformation
\begin{align}
     \mqty(a_\vk(t)\\a_\vk^*(t))=\mathfrak{K}_\vk(t)\mqty(\tilde{a}_\vk(t)\\\tilde{a}_\vk^*(t)),
\end{align}
with 
\begin{align}
    \mathfrak{K}_\vk(t)=\mathfrak{J}_\vk(t)\tilde{\mathfrak{J}}^{-1}_\vk(t)=\mqty(\kappa_\vk(t)&\lambda_\vk(t)\\\lambda_\vk^*(t)&\kappa_\vk^*(t)).
    \label{kappa}
\end{align}
Thus, the Fock representation defined by $\tilde{\mathfrak{J}}_\vk(t)$ will be unitarily equivalent to the reference representation if and only if the Bogoliubov transformation given by $\mathfrak{K}_\vk(t)$ can be implemented as a unitary operator in the Fock space defined by the first representation. This condition, as we have seen, imposes the integrability of the squared modulus of the $\lambda_\vk$ coefficients, which is given explicitly  by
\begin{align}
     \abs{\lambda_\vk(t)}=\abs{f_\vk(t)\tilde{g}_\vk(t)-g_\vk(t)\tilde{f}_\vk(t)},
\end{align}
as one can easily obtain from \eqref{kappa}.
But, since we previously required that both $\mathfrak{J}_\vk(t)$ and $\tilde{\mathfrak{J}}_\vk(t)$ must satisfy the criterion of unitary implementation of the dynamics, their coefficients will be of the form \eqref{generalform}, and hence
\begin{align}
    \abs{\lambda_\vk}=\frac{1}{2}\abs{\frac{\tilde{h}_\vk}{h_\vk}-\frac{{h}_\vk}{\tilde{h}_\vk}}.
\end{align}
Provided that both $h_\vk$ and $\tilde{h}_\vk$ have the behavior given in \eqref{h-cond}, we conclude that $\abs{\lambda_\vk}=\mathcal{O}(k^{-2})$ in the ultraviolet regime.
Thus, the integral of the squared modulus of these coefficients will converge, and both representations will be unitarily equivalent, as we wanted to show.

\section{Schwinger effect}
We now apply the ideas developed in the last sections to a particular example of particle creation due to Schwinger effect. To do so, we consider the following time dependent potential, known as Sauter-type potential \cite{Sauter1931742}: 
\begin{align}
    \vA(t)=(0,0,0,A(t)),\quad A(t)= \mathcal E\tau H_\tau(t),
    \label{potential}
\end{align}
where $H_\tau(t)=\frac12[\tanh(t/\tau)+1]$.
This corresponds to a pulse in the electric field, rapidly decreasing both in the asymptotic past and future, 
\begin{align}
    E(t)=-\dot{A}(t)=\frac{-\mathcal E}{2\cosh^2(t/\tau)},
    \label{electricfield}
\end{align}
where  $\mathcal E/2$ is the maximum amplitude  of the electric field, and $\tau$ is the characteristic timescale during which the pulse occurs.
 We set initial conditions in the asymptotic past, $t_0\rightarrow -\infty$, so that $A(t_0\rightarrow-\infty)\rightarrow0$ in agreement with our previous discussion. Note that for this potential, $\dot\omega/\omega$ is finite and therefore the asymptotic analysis of the dynamics carried out in section \ref{asymp} applies to this case. Furthermore, we have chosen this particular potential since it allows for an analytical solution of the equation of motion for each mode, as we will see.

Substituting \eqref{potential} into \eqref{modeqmotion} we obtain the following equation 
\begin{align}
   \Ddot{z}_\vk+\qty{[k_3+q\mathcal E\tau H_\tau(t)]^2+m^2+k^2_\perp}z_\vk=0,
   \label{particulareq}
\end{align}
being $k_3$ the component of the wave vector in the direction of the vector potential, and $k^2_\perp=k^2-k_3^2$.

We now search for (complex) solutions $\zeta_\vk$ of \eqref{particulareq} with asymptotic initial conditions such that each mode behaves as a plane wave satisfying the free Klein-Gordon equation in the asymptotic past. In other words, up to an irrelevant phase
\begin{equation}
    \zeta_\vk(t)\sim  e^{-i\omega_0t}\qquad \text{when}\qquad t\rightarrow -\infty.
  \label{asympcond}
\end{equation}

A solution of \eqref{particulareq} satisfying \eqref{asympcond} is given by \cite{Navarro}:

\begin{align}
    \zeta_\vk(t)=e^{-i\omega_0t}\qty(1+e^{2t/\tau})^{\tfrac{(1-i\delta)}{2}}{}_2F_1\qty(\rho_+,\rho_-;1-i\tau\omega_0;-e^{2t/\tau})
    \label{solution},
\end{align}
where ${}_2F_1$ is the ordinary hypergeometric function \cite{abramowitz1965handbook}, and $\delta,\rho_\pm$ are constants defined by
\begin{align}
    \delta=\sqrt{(q\mathcal E\tau^2)^2-1},\quad
  \rho_\pm=\frac{1}{2}\qty{1- i[\tau(\omega_0\pm \omega_\infty)+\delta]},
\end{align}
with $\omega_\infty\equiv\omega(t=\infty)=\sqrt{(k_3+q\mathcal E\tau)^2+m^2+k_\perp^2}$.

Substitution of this solution \eqref{solution} into \eqref{betasquare} gives an explicit expression for $|\beta_\vk(t)|^2$ as a function of time in terms of hypergeometric functions (indeed, since the derivative of a hypergeometric function is proportional to other hypergeometric function). Thus, using the asymptotic properties of these functions \cite{abramowitz1965handbook}, it is straightforward to check that in the asymptotic future $t\rightarrow\infty$, when the electric field vanishes, we recover the result of \cite{Gavrilov:1996pz,Navarro}:
\begin{equation}
|\beta_\vk(t\rightarrow \infty)|^2\rightarrow \frac{\cosh{[\pi\tau(\omega_\infty-\omega_0)] }+\cosh{(\pi\delta)}}{2\sinh{(\pi\tau\omega_0)}\sinh{(\pi\tau\omega_\infty)}},
\end{equation}
as one would expect. Furthermore, our approach allows for a well-defined particle number density ${N_\vk(t)=|\beta_\vk(t)|^2}$ at any instant of time.

In other words, thanks to the fact that in our Fock quantization the dynamics is implemented unitarily, we have a well defined concept of particle (and antiparticle) along the entire evolution, and  we can  therefore compute the total number of particles  plus antiparticles  created out of the vacuum at any instant of time,
\begin{equation}
    N(t)=\frac{1}{2}\int_{\mathbb{R}^3}d\vk^3\, (|\beta^a_\vk(t)|^2+|\beta^b_\vk(t)|^2)=\int_{\mathbb{R}^3}d\vk^3\,|\beta_\vk(t)|^2.
    \label{Nparticles}
\end{equation}                 
Here we have explicitly reminded that there are two contributions coming from the real and imaginary parts of the original complex scalar field respectively, but we choose $\mathfrak{J}_\vk^a(t)=\mathfrak{J}_\vk^b(t)$, so that these contributions  are identical and equal to \eqref{betasquare} so that charge conservation is not violated.
 The integral in \eqref{Nparticles} converges by construction, as that is precisely what we have demanded when choosing our complex structure.
 
\begin{figure}
  \centering
    \includegraphics[width=0.49\textwidth]{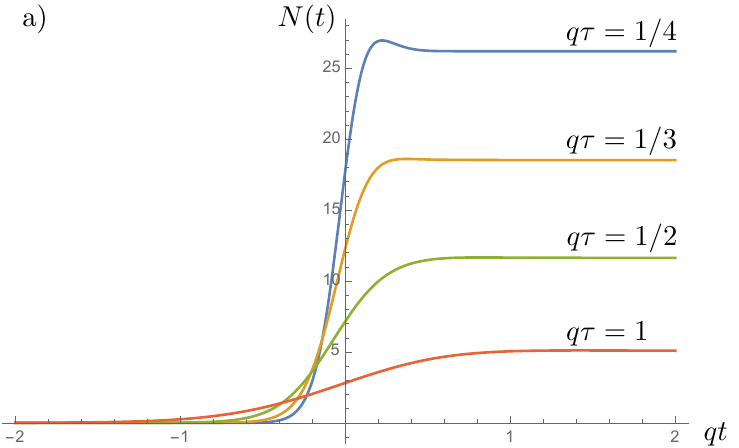}
    \hspace{0.05cm}
     \includegraphics[width=0.49\textwidth]{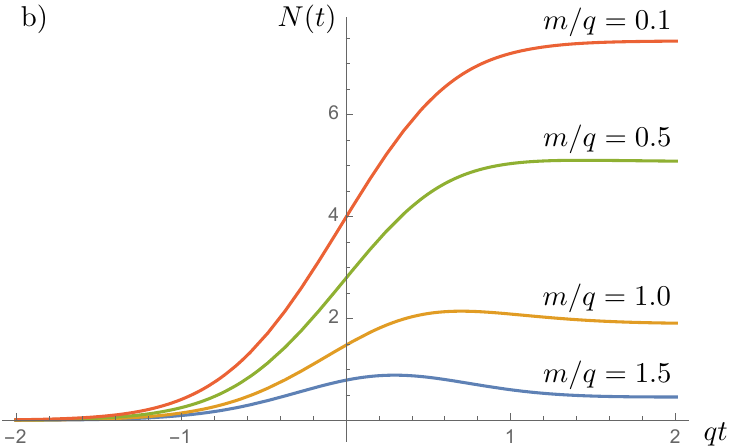}
     \caption{Evolution of the total number of particles plus antiparticles for $\mathcal E\tau=5$. a) Different values of $q\tau$ and $m/q=0.5$. b) Different values of $m$ and  $q\tau =1$.}
     \label{fig}
\end{figure}

 As an example, Figure \ref{fig} shows the evolution of $N(t)$ for the case $\mathcal E\tau=5$, and for different values of the mass $m$ of the scalar field and of the parameter $\tau$ that measures the adiabaticity of the electric pulse. To create these plots, for each value of $t$, the integral in \eqref{Nparticles} has been effectively computed with a high enough cut-off for the wave numbers $k_3$ and $k_\perp$ so that convergence is reached. Obviously $N(t)$ is finite and also evolves smoothly in time, converging to a final asymptotic value because the considered electric field rapidly decreases and vanishes asymptotically.
  Our results concerning the evolution of the particle number with time are very similar to those obtained in \cite{TANJI20091691}  for a slightly different external potential, and \cite{Huet2014} in the exact same case. We would like to note that the quantization chosen in \cite{TANJI20091691} is unitarily equivalent to the one that we have proposed in this work, so that it constitutes another example of the family of unitary equivalent complex structures which is selected via unitary implementation of the dynamics, although this is not explicitly justified in \cite{TANJI20091691}.
 Interestingly, for the range of parameters that we have investigated, we do not find that $N(t)$ oscillates (as it does in \cite{Huet2014}), although it reaches the correct asymptotic value when the external field vanish. However, it does not monotonically grow either in all cases. Indeed, for fixed mass, $N(t)$ develops a hump with a local maximum as the pulse becomes faster, and for fixed pulse, $N(t)$ also develops a similar hump as the mass of the field increases. This behavior is more similar to the results of \cite{TANJI20091691}. A satisfactory explanation of this fact would indeed require a physical interpretation of the quantity $N(t)$ at all times. If we interpret it not as the real particle number, but as an expectation value of the number of quanta at a given instant, this humps may be explained following the arguments given in \cite{Fedotov2011} in terms of vacuum fluctuations in the transient region.
 
  In any case, this $N(t)$ is one of many and not necessarily 
 the \emph{physical} particle number, understood as the expectation value of the number of quanta of the field that one would measure at a given time. As commented before, our $N(t)$ is just the consequence of our choice of quantization. Any other choice within our equivalence class would lead to another (equally well defined) time dependence of $N(t)$. 
 Our contribution to previous discussions on this issue (see e.g. \cite{Dabrowski_2016,Hebenstreit:2008ae,Kim:2011sf}) is that the unitary implementation of the evolution is a criterion that the \emph{physically correct} quantization (chosen according to additional criteria) should satisfy, so that we think that the correct \emph{physical} particle number must correspond to a quantization unitarily equivalent to the one that we present here.

\section{Conclusions}

In this work, we have come up with a procedure to deal with the ambiguities that appear in the canonical quantization of a charged scalar field coupled to a (classical, non-dynamical) gauge field, including the  ambiguities  due to the gauge freedom in the description of the external field. We have applied the criterion of unitary implementation of the dynamics to select a particular family of complex structures, which define a unique Fock representation up to unitary equivalence. Our work extends the approach of a long list of previous analyses done in gravitational homogeneous spacetimes (see \cite{Cortez:2016xsn,Cortez_2020,Cortez:2016oxm,Cortez:2016kwy, Cortez:2015mja, Cortez:2012cf,Gomar:2012xn} and references therein) to the case of the Schwinger effect. Although other approaches to the same problem in this context have been proposed in the literature, e.g. \cite{Dabrowski_2016},
we find that our criterion allows, in particular, for a well defined notion of the number of particles created by the external perturbation at all times in the evolution, and not only in the asymptotic regime.

More concretely,   we have  shown that once the gauge freedom is fixed in a specific way and the quantization of the field is carried out by means of a chosen complex structure $J$,  there exist a parametric family of complex structures $J^g$, with $g$ the elements of the gauge group, which define physically equivalent quantizations in any other gauge fixation. The gauge transformation is not implemented quantum-mechanically as an operator acting on the Fock space of the original quantization. Instead, it is dealt with as a possibly time-dependent classical rescaling. In this way, if the initial complex structure $J$ leads to a unitary implementation of the evolution, 
any other physically equivalent quantization defined by such $J^g$ will have the same quantum unitary dynamics up to time-dependent classical factors. 

One encounters a similar situation in cosmological settings, where one has for example the freedom to  canonically transform the field by means of a time-dependent global rescaling \cite{Cortez:2015mja}. We quantize the field using the most convenient  parametrization, which is the one that will eventually yield the Minkowski vacuum in the absence of external field after quantization, and deal with the global rescaling classically if later we want to change to a different parametrization.

In general, whenever one tries to perform the canonical quantization of a field,
both the freedom in the choice of complex structure and the field parametrization (or gauge choice) can be taken into account at once by letting the complex structure to depend on time. This is equivalent to splitting the time-dependence of the field when written as an expansion in terms of bases of functions that span the particle-like and antiparticle-like sectors of the field: some time-dependence is assigned to the elements of the bases (which are classical) and the rest to the annihilation and creation-like variables (which are later promoted to operators in Fock space).

We have explicitly shown that the joint criteria of requiring the complex structure to be invariant under the symmetries of the equations of motion, together with the condition that the quantum dynamics of the creation and annihilation operators to be unitarily implementable in Fock space, restrict quite a lot the form of the allowed complex structures. More importantly, all the complex structures fulfilling these conditions turn out to define a unique equivalence class of Fock representations, defining unitarily equivalent quantizations. In summary, we have eliminated the ambiguities in the quantization of a scalar field in the presence of an arbitrary, spatially-homogeneous and time-dependent external electric field   up to unitary equivalence between quantizations.

The uniqueness proof presented in this work strongly relies on the asymptotic  ultraviolet  behavior of the solutions of the Klein-Gordon equation with the electric field written in the temporal gauge, in which the solutions can be written in terms of their Fourier components. 
By analyzing the most general form that an invariant complex structure may have, we have seen that the nontrivial unitary implementability of the evolution requires a very specific asymptotic behavior for the annihilation and creation-like variables, encoded in the asymptotic behavior of the complex-structure coefficients, given in \eqref{generalform}-\eqref{h-cond} (up to a phase, according to \eqref{conditionsuma}).  
This asymptotic behavior of the complex structures is in fact needed for the unitarity of the dynamics, and we have shown that all complex structures which have this asymptotic behavior will be unitarily equivalent. 

Our method applies to a very general class of electric fields, and not just to those that vanish both in the asymptotic past and future (i.e. field configurations localized in time).
It is worth mentioning that the complex structure that defines the Minkowski vacuum in the absence of external electromagnetic field, is \emph{not} included within the equivalence class of quantizations that allow a unitary dynamics unless the electromagnetic field vanishes. Hence, the unitary implementation of the evolution in the Fock space associated to the Minkowski-like vacuum (i.e. without external electromagnetic field) will not be possible for arbitrary, time-dependent vector potentials, which is in fact the result previously obtained by Ruijsenaars in \cite{Ruijsencharged}. 

On the other hand, for external fields which vanish both in the asymptotic past and future, the asymptotic analysis of the particle creation taking into account our reference complex structure  is  equivalent to that obtained via the   complex structure in Minkowski, since both complex structures coincide in the limit of vanishing external field. In other words,  the complex structure defining the Minkowski vacuum  is perfectly fine to analyze the asymptotic particle creation spectrum for external electromagnetic fields which vanish in the asymptotic past and future--- see \cite{Gavrilov:1996pz} for some analytical examples of this kind of systems---.   We have in particular considered a Sauter-like solvable potential. Not only we recover the particle creation in the asymptotic future that was obtained in previous literature, but we also show its behavior at any instant of time, and compare it with the one obtained previously in the literature by means of other approaches.

\acknowledgments
Financial support was provided by the Spanish Government through the projects  FIS2017-86497-C2-2-P (with FEDER contribution) and FIS2016-78859-P (AEI/FEDER,UE).
AGMC acknowledges financial support from the Ministry of Education,
Culture, and Sports, Spain (Grant No. FPA2017-83814-P), the Xunta de
Galicia (Grant No. INCITE09.296.035PR and Conselleria de Educacion), the
Spanish Consolider-Ingenio 2010 Programme CPAN (CSD2007-00042), Maria de
Maetzu Unit of Excellence MDM-2016-0692, and FEDER.

\bibliographystyle{JHEP}
\bibliography{biblio}
\end{document}